\documentclass[12pt,preprint]{aastex}
\def\spose#1{\hbox to 0pt{#1\hss}}
\def\lta{\mathrel{\spose{\lower 3pt\hbox{$\mathchar"218$}} \raise 2.0pt\hbox{$\mathchar"13C$}}}
\def\gta{\mathrel{\spose{\lower 3pt\hbox{$\mathchar"218$}} \raise 2.0pt\hbox{$\mathchar"13E$}}}


\lefthead{Ramirez-Ruiz \& Madau}
\righthead{Unveiling misaligned GRB jets}
\begin{document}

\title{Compton Echoes from Gamma-Ray Bursts: Unveiling Misaligned Jets in 
Nearby Type Ib/c Supernovae}

\author{Enrico Ramirez-Ruiz\footnote{Chandra Fellow.}} 
\affil{School of Natural Sciences, Institute for Advanced Study, Einstein 
Drive, Princeton, NJ 08540; enrico@ias.edu}
\and 
\author{Piero Madau}
\affil{Department of Astronomy \& Astrophysics, University of California, 
Santa Cruz, CA 95064; pmadau@ucolick.org}

\begin{abstract}
There is now compelling evidence of a link between long-duration
gamma-ray bursts (GRBs) and Type Ib/c supernovae (SNe). These
core-collapse explosions are conjectured to radiate an anisotropic,
beamed component associated with a decelerating, relativistic outflow
and an unbeamed, isotropic component associated with the slowly
expanding stellar debris.  The anisotropic emission remains at a very
low level until the Doppler cone of the beam intersects the observer's
line of sight, making off-axis GRB jets directly detectable only at
long wavelengths and late times ($\gta 1$ yr). Circumstellar material,
however, will Compton scatter the prompt gamma-ray and afterglow
radiation flux and give rise to a reflection echo. We show that the
Compton echo of a misaligned GRB carries an X-ray luminosity that may
exceed by many orders of magnitude that produced by the underlying
subrelativistic SN during the first few weeks. Bright scattering
echoes may therefore provide a means for detecting a population of
misaligned GRBs associated with nearby Type Ib/c SNe and yield crucial
information on the environment surrounding a massive star at the time
of its death. The question of whether the interpretation of GRB980425
as an ordinary GRB observed off-axis is consistent with the lack of an
X-ray echo is addressed, along with the constraints derived on the
possible existence of misaligned GRB jets in SN1993J, SN1994I,
SN1999em, and SN2002ap.
\end{abstract}

\section{Introduction}

The first hint of a connection between core-collapse supernovae (SNe)
and gamma-ray bursts (GRBs) came in April of 1998, when SN1998bw was
detected in the {\it BeppoSAX} error box of GRB980425 (Galama et
al. 1998). While unremarkable in its apparent properties (peak flux,
duration, burst profile), GRB980425 (isotropic) energy release of only
$8\times 10^{47}$ erg was some four orders of magnitude smaller than
that of {\it typical} GRBs (e.g., Bloom et al. 1998). The observational
basis for a connection between GRBs and Type Ib/c SNe has been greatly
strengthened by the recent discovery (Hjorth et al. 2003; Stanek et
al. 2003) of a SN1998bw-like spectrum lurking beneath the optical
afterglow of GRB030329. This SN has been designated SN2003dh. Due to
its extreme brightness and slow decay, spectroscopic observations of
GRB030329 have been extensive. The early spectra consisted of a
power-law decay continuum ($F_\nu \propto \nu^{-0.9}$) typical of GRB
afterglows with narrow emission features identifiable as H$\alpha$,
[OIII], H$\beta$ and [OII] at $z=0.1687$ (Greiner et al. 2003), making
GRB030329 the second nearest burst overall (GRB980425 being possibly
the nearest at $z=0.0085$) and the {\it classical} burst with the
lowest known redshift. At this distance, GRB030329 is typical in its
gamma-ray budget -- its total isotropic energy release is $9 \times
10^{51}$ erg (Hjorth et al. 2003).

Independent of their connection with a GRB, SN1998bw and SN2003dh
belong to the unusual class of Type Ic SNe. The lack of hydrogen lines
in both spectra is consistent with model expectations that the stellar
progenitor lost its hydrogen envelope to become a Wolf-Rayet (WR) star
before exploding (Woosley 1993). The broad lines
are also suggestive of an asymmetric explosion viewed along the axis
of most rapid expansion (e.g., MacFadyen \& Woosley 1999). The
radiated energy may therefore be considerably smaller than the total
explosive energy -- in the range $1-2 \times 10^{52}$ erg -- inferred
assuming spherical symmetry (Woosley et al. 1999). Moreover, the
larger ratio of afterglow luminosity to SN luminosity in GRB030329 is
consistent with a burst that has been observed nearly on axis, whereas
the low luminosity of GRB980425 and its afterglow hint to a
substantially off-axis view (Nakamura 1999; Eichler \& Levinson 1999;
Woosley et al. 1999; Granot et al. 2002). In these models, together
with the afterglow radiation restricted to a narrow relativistic beam,
there is also quasi-spherical emission from slower moving SN ejecta
running into the circumburst medium (Ramirez-Ruiz et al. 2002; Zhang
et al. 2004). The flux associated with the beamed component declines
rapidly with observer direction, thus making the task of detecting
misaligned GRB jets particularly challenging (Granot et al. 2002).

Even if GRBs were caused by a standard set of objects, their
appearance would depend drastically on orientation relative to the
line of sight. The gamma-rays we receive come from material the motion
of which is directed preferentially along the rotation axis of the
stellar progenitor, and therefore provide no information about the
ejecta in other directions. The propagation of gamma-rays through an
external medium would, however, randomize the initial photon pulse
energy and, because of light-travel time effects, produce a luminous
reflection echo (Dermer et al 1991; Madau et al. 2000; Ramirez-Ruiz et
al. 2001; Sazonov \& Sunyaev 2003). We show below that the X-ray
luminosity of such an echo exceeds during the first few weeks that
produced by the underlying subrelativistic SN. The detection of
energetic scattered light, with luminosities high enough to be
detected by {\it Chandra} and {\it XMM-Newton} observations (and the
{\it Swift} experiment now under construction), may therefore provide
a means for uncovering a population of misaligned GRB jets associated
with nearby Type Ib/c SNe. Conversely, one may use the lack of
evidence for a Compton echo to set constraints on the the density
profile of the circumburst material that surrounds a massive star at
the time of its death.

\section{X-ray Emission from Misaligned GRB Jets}

\subsection{Off-axis Afterglow Emission}

The deceleration of a relativistic jet propagating through a stellar
wind starts when about half the initial energy has been transferred to
the shocked matter, i.e. when an amount $\Gamma_0^{-1}$ of its own
rest mass has been swept up. The relativistic expansion is then
gradually slowed down, and the blast wave evolves in a self-similar
manner with a power-law lightcurve (Blandford \& McKee 1976). This
phase ends when a total mass $E_\Omega/c^2$ (we denote with $E_\Omega$
the burst energy per unit solid angle) shares the energy and the
Lorentz factor drops to 1. As, in the unshocked
stellar wind, the mass within radius $r$ is $\dot{M}r/v_w$, the blast
wave deceleration radius is $r_{\rm dec}=E_\Omega v_w /(
\dot{M}c^2\Gamma_0^2) \approx ( 2 \times 10^{15}\, {\rm cm})\,
E_{\Omega,53}v_{w,3} \dot{M}_{-5}^{-1} \Gamma_{0,2}^{-2}$, where
$\dot{M}_{-5}=(\dot{M}/10^{-5}\,{\rm M_\odot\,yr^{-1}})$ and
$v_{w,3}=(v_w/ 10^{3}\,{\rm km\,s^{-1}})$. At observer times of more
than a day, the blast wave would be decelerated to a moderate Lorentz
factor, irrespective of the initial value (Rhoads 1999). The beaming
and aberration effects are thereafter less extreme, so we observe
afterglow emission not just from material moving almost directly
towards us, but from a wider range of angles. A temporal break occurs
when the angle $\theta \sim \Gamma^{-1}$ of the Doppler beaming cone
equals the jet opening angle $\theta_0$. For a blast-wave decelerating
according to the approximation $\Gamma(r) = \Gamma_0$ for $r < r_{\rm
dec}$, and $\Gamma(r) = \Gamma_0 (r/r_{\rm dec})^{-s}$ for $r > r_{\rm
dec}$, the time $t_b$ at which the temporal break is observed after
the explosion is given by $t_b \approx (1+z)[- r\cos \theta_0 +
\int_0^r dr^\prime \beta^{-1}(r)]/c$. Here $c\beta(r)=
c[1-\Gamma(r)^{-2}]^{1/2}$ is the speed of the blast wave.  This gives
\begin{equation} 
t_b= r_{\rm dec}\;[(\Gamma_0\theta_0)^{1/s} (1-\cos\theta_0) + {2s +
(\Gamma_0\theta_0)^{2+1/s}\over 2 (2s+1)\Gamma_0^2} ] {(1+z) \over
c}\; \;.
\label{ttime}
\end{equation} 
The lightcurve seen by an observer located within the initial jet
aperture $\theta < \theta_0$ is very similar to that for an on-axis
observer (e.g., Dermer et al. 2000) but differs markedly in intensity
once the observer is outside the beaming cone. The emission remains at
a very low level until the Doppler cone of the beam intersects the
observer's line of sight.  This can be seen by comparing the
$\theta_0$ and $2\theta_0$ curves in Figure 1. Following the arguments
given to derive equation (\ref{ttime}), we find that an off-axis
observer ($\theta >\theta_0$) starts to detect emission at a level
comparable to an on-axis observer at times given by equation
(\ref{ttime}), but with $\theta_0$ replaced by $\theta-\theta_0$.  As
shown in Figure 1, it appears that the X-ray emission from highly
misaligned GRB jets (i.e. $\theta >3\theta_0$) is not significantly
larger than that emitted by the SN remnant itself.  Since the jet
opening angle of a GRB is typically assumed to be of the order of
$\theta_0 \le 10^\circ$ (Frail et al. 2001; which represents 0.76\% of
the full sky) it is fair to conclude that detection of short
wavelength transients associated with core-collapse SNe will not
unambiguously demonstrate the presence of a GRB jet. The ratios of the
on-axis to off-axis fluxes are however much smaller at radio
frequencies where the emission persists at comparable levels for about
the same length of time (Paczy\'nski 2001). 

\subsection{Reflection Echoes} 
The beaming angle for the gamma-ray emission could be far smaller than
that of the late-time afterglow, and is much harder to constrain
directly. If the gamma rays were much more narrowly beamed than the
optical afterglow there should be many {\it orphan} afterglows. The
sudden brilliance of a GRB propagating in a preburst stellar wind will
be reflected by the circumstellar gas. If $E = \int E_{\epsilon
\Omega} d\epsilon d\Omega$ is the total energy emitted by the burst,
where $E_{\epsilon \Omega}$ is the energy emitted per unit energy
$\epsilon$ and unit solid angle $\Omega$, then the equivalent
isotropic luminosity of the Compton echo inferred by a distant
observer is
\begin{equation} L_{\epsilon'}=4\pi\int n_w(r)
E_{\epsilon\Omega} {d\sigma\over d\Omega} {dr \over dt}d\Omega,
\label{lumecho} \end{equation} 
where $r$ is the distance from the site of the burst, $d\sigma /
d\Omega$ is the differential Klein-Nishina cross section for
unpolarized incident radiation, $\epsilon'=\epsilon[1+\epsilon
\,(1-\cos \varphi)/m_ec^2]^{-1}$ is the energy of the scattered
photon, and $n_w$ is the local electron density (Madau et al. 2000).
The equal-arrival time scattering material lies on the paraboloid
determined by $r=ct (1-\cos \varphi)^{-1}/(1+z)$, where $\varphi$ is
the angle between the line of sight and the direction of the
reflecting gas as seen by the burst (e.g., Blandford \& Rees 1972),
$t$ is measured since the burst is first detected, and $(1+z)$ is the
cosmological time dilation factor for a source at redshift $z$. At
energies below 100 keV, where the echo mirrors the spectral energy
distribution of the prompt pulse, echoes will typically be harder than
the X-ray afterglows observed at around 1 day. Above 200 keV, and
independently of the incident gamma-ray spectrum, reflection echoes
will show a high-energy cutoff because of Compton downscattering
(Madau et al. 2000). One distinctive feature of the echo signal is
that the scattered radiation is likely to be polarized. Even when the
incident beam is unpolarized, the expected fractional polarization is
$\Pi=(1-\cos^2\varphi)/(1+\cos^2\varphi)$. Therefore, in the case of
observing the jet precisely at $\varphi=90^{\circ}$, the scattered
radiation is 100\% polarized. On the other hand, if the prompt
radiation is 100\% polarized (e.g., Coburn \& Boggs 2003), no net
intensity will be seen at $\varphi=90^{\circ}$, since the electron's
motion is confined to a plane normal to the observer\footnote{The
differential cross sections for 100\% polarized and unpolarized
radiation, are in the ratio $2\cos^2\varphi\;:\;1+\cos^2\varphi$
(e.g., Rybicki \& Lightman 1979)}.

The reflected flash luminosity of a collimated GRB propagating through
a constant velocity wind with $n_w(r)=Ar^{-2}$ is dominated by
the receding jet as, at a given observer time, it originates closer to
the GRB where the density is higher. In this case,
\begin{equation}
\epsilon L_\epsilon={3 A \epsilon E_\epsilon \sigma_T\over 
4ct^2}\,\psi(\varphi)= (3\times 10^{41} {\rm erg\;s^{-1}})~
\psi(\varphi) \dot{M}_{-5} v_{w,3}^{-1} E_{51}\, (1+z)^2\, t_{\rm days}^{-2},
\label{lumecho2}
\end{equation}
(Madau et al. 2000), where $\sigma_T$ is the Thomson cross section
$E_{51}=E/ 10^{51} {\rm erg}$ is total energy of the collimated GRB
pulse, and $\psi(\varphi)\equiv (1+\cos^2 \varphi)$.  The echo
declines with time faster than most afterglows observed on-axis but
exceeds them in intensity once the observer is outside the beaming
cone (Fig. 1). For this reason, backscattered radiation could provide
a means for detecting a population of misaligned GRBs in nearby
core-collapse SNe and provide important information about the
circumburst environment.

Equations (\ref{lumecho}) and (\ref{lumecho2}) assume that photons
scatter only once and that absorption is negligible. As the reflecting
zone propagates along the receding beam, the backscattered radiation
on the way to the observer will pass with impact parameter $b=r\sin
\varphi$ through regions of even higher density than the scattering
layer.  For a jet inclined at angle $\varphi$ to the line of sight,
the electron column density traversed by the backscattered radiation
in the steady, spherically symmetric wind of a WR star is
$N_w=A(\pi-\varphi)/b \approx (6\times 10^{19}\,{\rm
cm}^{-2})\,\Psi(\varphi) \,\dot M_{-5}\,v_{w,3}^{-1}
\,(1+z)\,t^{-1}_{\rm days}$, where $\Psi(\varphi)\equiv
(1+\cos\varphi)(\pi-\varphi)/\sin\varphi$. This amounts to an optical
depth for Thomson scattering of $\tau_T \sim [10^{-2},2\times
10^{-4}]\,t_{\rm days}^{-1}(1+z)$ for $\varphi=[1^\circ,45^\circ]$.
Photoabsorption at 1 keV takes place with a cross section of $\sim
2\times 10^{-22}\,$ cm$^2$ (for solar abundances).  Multiple
scattering and photoabsorption can therefore be neglected all
times\footnote{The optical depth to electron scattering could in
principle be much higher if some lines of sight to the observer
traversed the SN ejecta. This is unlikely, however, as it requires
$v_{\rm SN}/c > \sin\varphi/(1+\cos\varphi)$, corresponding to $v_{\rm
SN}/c>0.2$ for jets with $\varphi >20^\circ$, where $v_{\rm SN}$ is
the velocity of the SN ejecta.}  but in the very early light curve.

\section{A GRB Behind the Veil?}

Collimated GRB outflows directed away from an observer will produce
extended X-ray transients at a flux level orders of magnitude lower
than transients observed along the jet axis. The off-axis jet
interpretation of GRB980425 requires the viewing angle to have been
$\ge 3\theta_0$ with $\theta_0$ the half angle of the most energetic
part of the jet, otherwise the X-ray afterglow would have contaminated
the SN light curve unacceptably (Fig. 1; similar requirements have
been derived for the optical light curve by Granot et al. 2002). The
off-axis jet interpretation of GRB980425 typically assumes that our
line of sight is a few degrees from the sharp-edge conical jet (see
dashed lines in Fig. 1). This interpretation would be inconsistent
with radio observations since in this case the jet deceleration would
bring our line of sight into the jet's radiation beaming cone, leading
to strong emission on timescales $<1$ year (Waxman 2004a). Our line of
sight must make a large angle with the jet axis ($\ge 4\theta$) in
order to avoid observing strong radio emission. The observed gamma-ray
flux of GRB980425 can be explained in this case by assuming that the
jet is not sharp-edged, but rather has wings of lower Lorentz factors
that extend to large $\theta$ (Woosley et al. 1999) and produce the
observed faint burst. Interestingly, this material would also increase
the afterglow luminosity at early times when almost nothing is seen of
the inner jet. At early times, the X-ray emission of the underlying
subrelativistic SN is probably small (Waxman 2004b). The X-ray
emission during the first few weeks should be therefore dominated by
the rapidly fading echo (Fig. 1). The fact that the echo decay is both
so steep and presents little spectral evolution will unambiguously
distinguish it from any relativistic blast wave in which the electrons
emit a constant fraction of the energy gained in the shock (Cohen et
al. 1998). The lack of a detectable Compton echo therefore places
severe constraints on the brightness of a possible GRB jet associated
with SN1998bw, and it is to this problem that we now turn our
attention.

During the first day after the explosion no significant flux was
observed by {\it BeppoSax} in the direction of GRB980425 to a limit of
about $10^{41}$ erg s$^{-1}$. An off-axis jet with standard $10^{51}$
erg energy expanding into a wind with $\dot{M} \approx 10^{-5}{\rm
M_\odot\,yr^{-1}}$ and $v_w\approx 1000\,{\rm km\,s^{-1}}$ can
therefore be ruled out. A misaligned jet interpretation for GRB980425
may still be consistent with the observations if it is assumed that
the wind of the progenitor was atypically-low density, with $\dot
M_{-5}/v_{w,3} < 0.1$ (Fig. 2). The lower value of
$\dot{M}_{-5}/v_{w,3}$ is consistent with the observed radio emission
from the shock driven into the progenitor wind by the SN ejecta
(Waxman \& Loeb 1999; Li \& Chevalier 1999), although, as usually in
the case of radio SNe, the data are unable to brake the degeneracy
between $\dot{M}/v_{w}$ and $\epsilon_B$ (i.e. the fraction of the SN
post-shock thermal energy carried by magnetic field). A model with
$\epsilon_B \ll 1$ and $\dot{M}_{-5}/v_{w,3} > 0.1$ would need
$\epsilon_B \le 10^{-4}$ in order to reconcile radio observations with
the prediction of an off-axis jet (Waxman 2004a; Soderberg et
al. 2004). If, on the other hand, one adopts a model with $\epsilon_B$
near equipartition, then $\dot{M}_{-5}/v_{w,3} < 0.1$ is required. Our
analysis demonstrates that the lack of detection of a GRB echo places
robust constraints -- i.e. independent of $\epsilon_B$ and
complementary to those derived via radio observations -- on the
assumed density profile.  By the same token, similar constraints on
the intrinsic energy of a misaligned GRB, compatible with the
circumstellar density profile inferred from radio observations, can be
placed for SN1993J, SN1999em, SN2002ap and SN1994I (Fig. 2).

\section{Summary} 

GRB explosions are conjectured to radiate an anisotropic, beamed
component from the decelerating, ultrarelativistic outflow and an
unbeamed, isotropic component from the slowly expanding stellar
debris.  The flux associated with the former depends upon the observer
direction and declines rapidly with observer time (Fig.  1). Hence the
X-ray flux detected by an on-axis observer is much larger than the
flux that would be detected from an identical source viewed off-axis.
The relative flux levels of on-axis and misaligned jet sources are,
however, much smaller in the radio/infrared than at optical and X-ray
frequencies.  Transients from misaligned GRB outflows could then be
detected at late times with radio follow-up observations (e.g., Granot
\& Loeb 2003). Here we have considered an alternative possibility: the
detection of scattered radiation from a narrowly collimated but
misaligned pulse of gamma-rays propagating through a WR-type
wind. While emission from the SN shock would dominate over the
scattered radiation at late times because of the steeper lightcurve of
a reflection echo, this will not be true at early times when the echo
may be 3 orders of magnitude brighter. Compton echoes in nearby SNe
could be studied with {\it Chandra} and {\it XMM-Newton} observatories
out to a distance of 0.1 Gpc. The planned {\it Swift} experiment may
also soon provide some constraints for individual bursts at higher
threshold energies.

We have argued that the lack of a detectable echo in GRB980425 may be
consistent with an off-axis jet interpretation only if the density of
the circumburst wind is $\dot{M}_{-5}/v_{w,3}<0.1$, lower than
expected for a WR star. The mass-loss history of a WR star during its
last few centuries could be quite complicated as the star enters
advanced burning stages unlike those in any WR star observed so
far. The lightcurve of the resulting echo will depend fairly strongly
on the properties of the progenitor system, especially the mass-loss
rate of the star. The deceleration of a presupernova wind by the
pressure of the surrounding medium could create circumstellar shells,
as would the interaction of fast and slow winds from massive
stars. The complex density structure seen in SN1987A may be a hint of
the immediate neighborhood of a GRB. Further data on X-ray emission
from nearby core-collapse SNe will therefore offer important clues to
the nature of the precursor star.

\begin{acknowledgements} 
We have benefited from many useful discussions with J. Granot,
C. Kouveliotou, B. Paczy\'nski, M. Rees, and E. Waxman. We are
particularly grateful to S. Woosley and the anonymous referee for
helpful insights regarding the calculations. This work is supported by
NSF grant AST-0205738 (PM), and NASA through grant NAG5-11513 (PM) and
a Chandra Postdoctoral Fellowship award PF3-40028 (ERR). Part of this
work was done while ERR was visiting the UCSC.
\end{acknowledgements}

\begin{figure} 
\caption{\footnotesize X-ray emission from misaligned GRB
jets. Compilation of observed X-ray lightcurves for GRBs and
core-collapse SNe (Kouveliotou et al. 2004) presented in the source
frame using a cosmology with $\Omega_m=0.27$, $\Omega_\Lambda=0.73$,
and $H_0=72\;{\rm km\;s^{-1}\;Mpc^{-1}}$. {\it Dashed lines:}
Lightcurves calculated at various inclination angles $\theta$ for a
GRB with standard parameters: $E_{\Omega,54}=1$, $\Gamma_0=100$,
$p=2.5$, $\epsilon_e=0.5$, $\epsilon_B=10^{-3}$ and $\theta_0=5^\circ$
(where $\epsilon_B$ and $\epsilon_e$ are the fraction of the internal
energy in the magnetic field and electrons, respectively, and $p$ is
the power-law index of the electron energy distribution). {\it Solid
lines:} Compton echo of a GRB. Lower two curves assume $L = L_0\;
t_{\rm days}^{-2}$ with $L_0 =10^{40}$ and $10^{42}$ erg s$^{-1}$ (see
eq. [\ref{lumecho2}]). The primary burst is assumed to be a two-sided
collimated pulse propagating at an angle $\varphi \sim 1$ rad and
invisible to the off-axis observer. The upper curve (highest
luminosity echo) is calculated for the dense environment expected at
the end of the evolution of a 29 ${M}_\odot$ WR star with solar
metallicity (Garcia-Segura et al. 1996). The collimated GRB, whose
total energy is $10^{52}$ erg, is assumed to have a broken power-law
spectrum, $\epsilon E_\epsilon \propto \epsilon$ if $\epsilon \le
\;250\;{\rm\;keV}$ and $\epsilon^{-0.25}$ otherwise. It is instructive
to look at the contribution of the red-supergiant shell (of radius
$r_{\rm sh}$) to the luminosity of the scattered echo. The time
interval, 2$r_{\rm sh} \cos \varphi /c$, between the two reflection
spikes at times $\sim 1000-2000$ days can be used to determine
$\varphi$.}
\label{fig1} 
\end{figure} 

\begin{figure}
\caption{\footnotesize Constraints on the density of circumstellar
material and burst energetics.  The lack of an X-ray echo at about 1
day from the explosion poses severe constraints on the existence of a
misaligned GRB jet in SN1998bw (shaded region).  An off-axis jet
interpretation may be consistent if the density of the stellar medium
is lower than expected for the assumed WR progenitor. Constraints on
the total kinetic energy derived from modeling of the radio emission
of SN1998bw are shown as open circles (Waxman 2004a; Li \& Chevalier
1999). Similar constraints on the possible existence of a misaligned
GRB jet compatible with the circumstellar density profile inferred
from radio observations (Weiler et al. 2002; black symbols) for
various core-collapse supernovae are shown. Constraints derived by
Panaitescu \& Kumar (2001) from the modeling the GRB afterglow are
also illustrated.}
\label{fig2} 
\end{figure}

\plotone{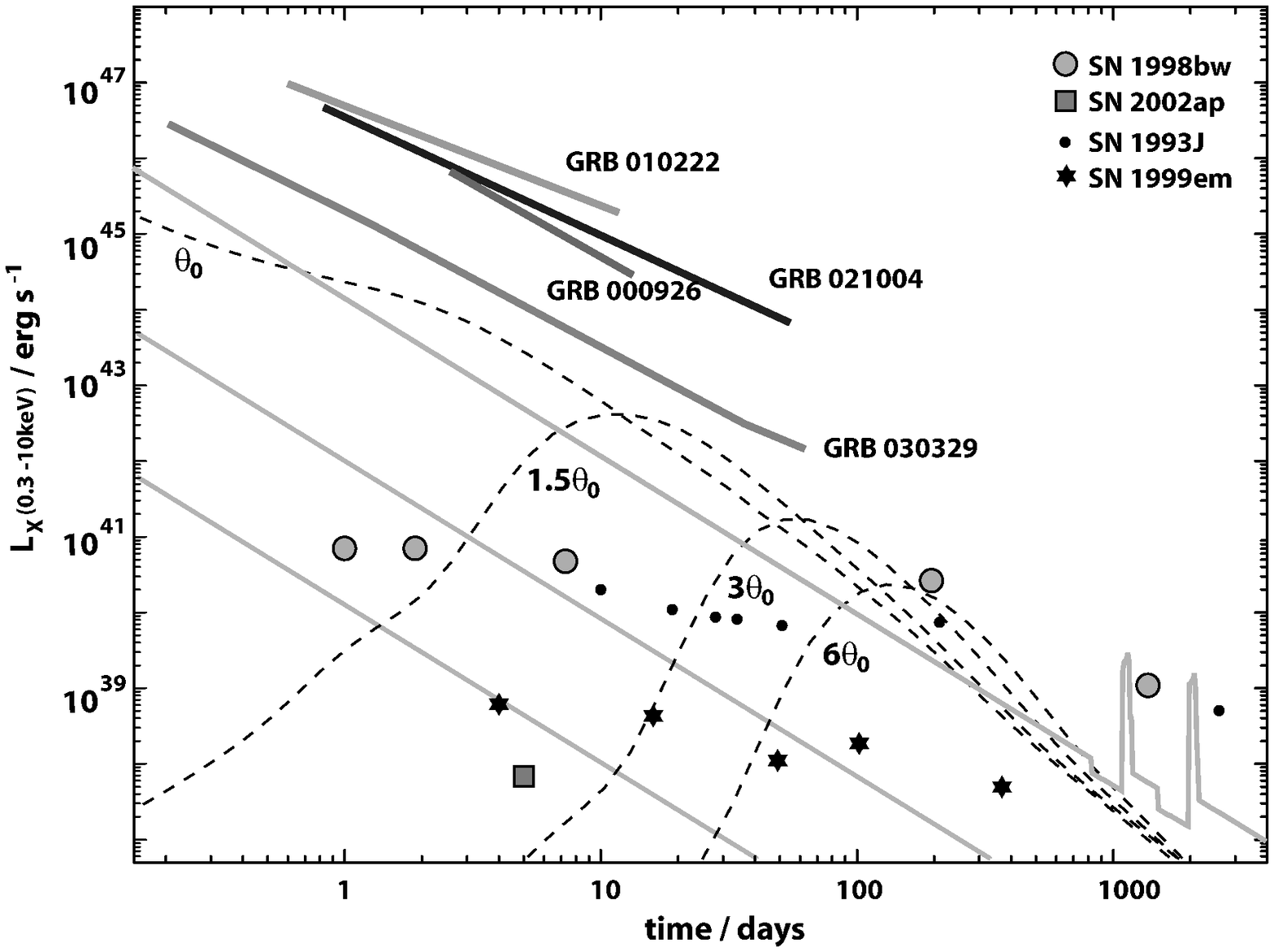}
\plotone{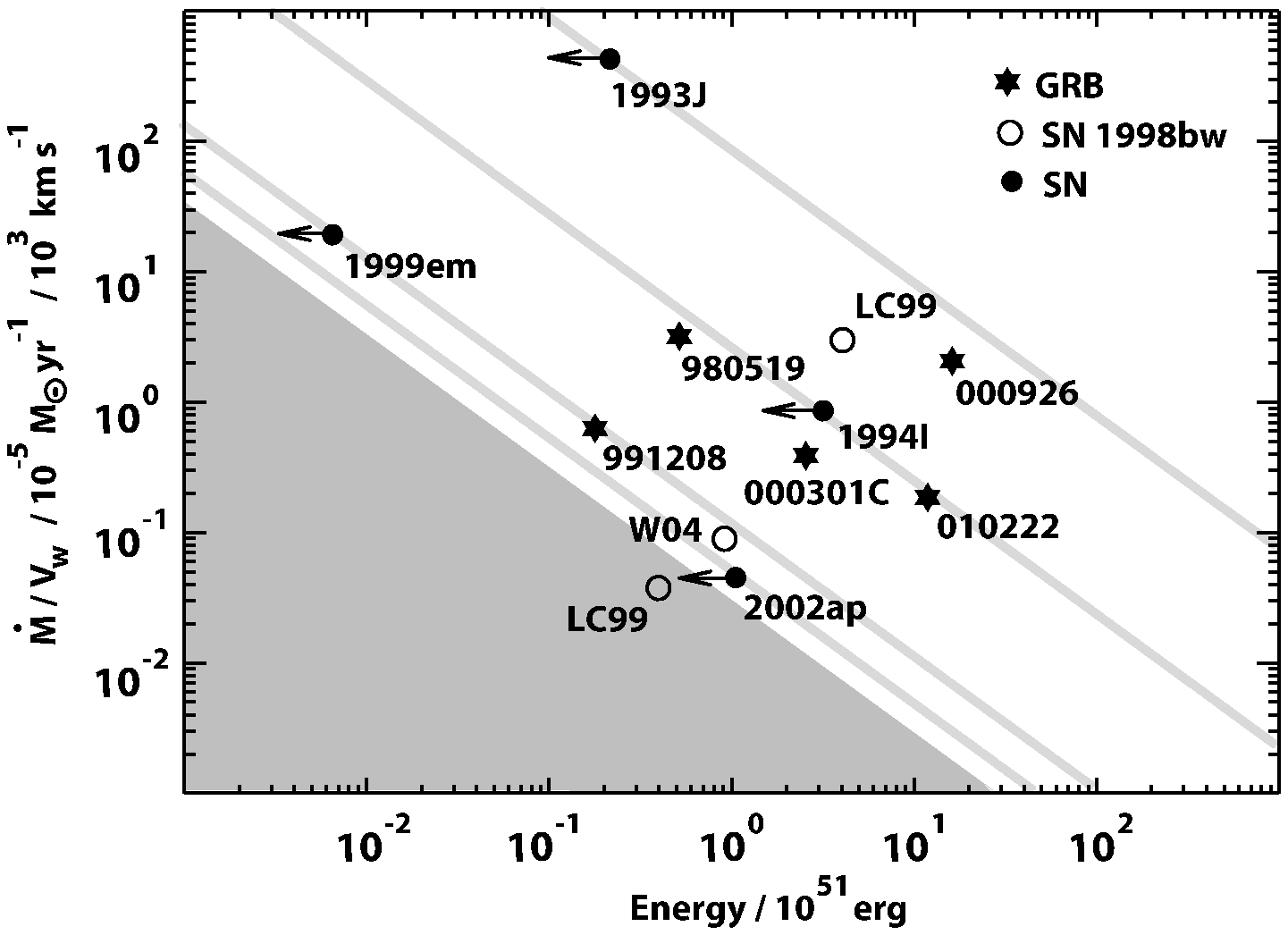}

\end{document}